-------------------------------------------------------------------
\documentstyle[aps,preprint]{revtex}
\begin{document}

\title{\bf Giant Josephson current through a single bound state in a
superconducting tunnel junction}
\author{ G. Wendin $^a$ and V. S. Shumeiko $^{a,b}$}
\address{
$^a$ Department of Applied Physics, Chalmers University of
Technology and
G\"{o}teborg University, S-41296 G\"{o}teborg, Sweden \\
$^b$ B.Verkin Institute for Low temperature Physics and
Engineering,
47 Lenin Ave., 310164
Kharkov, Ukraine}
\maketitle

\begin{abstract}
We study the microscopic structure of the Josephson current in a
single-mode tunnel
junction with a wide quasiclassical tunnel barrier. In such a
junction each
Andreev bound state carries a current of magnitude proportional to
the
{\em amplitude} of the normal electron transmission through the
junction.
Tremendous enhancement of the bound state current is caused by
the
resonance coupling
of superconducting bound states at both superconductor-insulator
interfaces
of the junction.
The possibility of experimental observation of the single bound
state current is discussed.
\end{abstract}

The Josephson effect in a tunnel junction deals with coherent
transmission
of Cooper pairs through a tunnel barrier which separates
superconducting electrodes.
What is the mechanism of such a transmission?
Conventional theory of the Josephson effect \cite{Jo,AB}, based on
the
phenomenology of the transfer Hamiltonian model \cite{Bardeen},
does not
provide adequate physical description of this process, e.g.
similar
to the quantum mechanical picture of tunneling of normal electrons.
Instead, it treats tunneling as a perturbative
transition, introducing a matrix element of coupling of electrons in
different electrodes
proportional to the amplitude of single electron tunneling \cite{Har}.

A more realistic description of Josephson tunneling,
based on the Bogoliubov-de Gennes equation \cite{dG}, was
suggested by
Furusaki and Tsukada \cite{Fu}. The crucial role in this picture is
played by superconducting bound states, similar to Andreev bound
states in SNS junctions \cite{Kk}. A bulk supercurrent,
when approaching the tunnel interface, experiences
transformation into current flowing through
superconducting bound states which provide transmission of
Cooper pairs through  the barrier.
The bound states are induced in the vicinity of the
junction by the discontinuity of the superconducting phase, and
they appear as a consequence of the current \cite{Ku}.
In a quantum junction the bound state
spectrum consists of a single pair of levels per transverse mode,
with symmetric position of the levels with respect to the chemical
potential. The Josephson current is distributed among bound states
in such a way that each bound state carries a current proportional to
the normal electron transparency of the junction.

In this Letter we show that the above picture of quantization of the
Josephson coupling is valid only for an extremely narrow barrier,
and that the picture is qualitatively different for any realistic
tunnel
barrier with large width on an
atomic scale. In the latter
case, the structure of the bound state spectrum is determined by
the coupling of superconducting surface states situated at the two
SI interfaces
of the SIS junction. In a symmetric junction, the
resonance coupling of these surface states provides tremendous
enhancement of the current flowing through a single bound state,
the magnitude being proportional to the {\em amplitude} rather
than the probability of normal electron tunneling. The currents
are distributed among the bound states in such a way that they
almost cancel each other in equilibrium, giving rise to a
comparatively
small residual current, including the  contribution from the
continuum.
This current coincides with the conventional Josephson
current given by Ambegaokar-Baratoff theory \cite{AB}. The large
current of the
single bound state can be revealed under nonequilibrium conditions
when the
bound level population is imbalanced by means of microwave
pumping or tunnel
injection.

We consider for the sake of clarity a single mode
quantum constriction with a rectangular potential barrier of length
$L$ and height $V$ (Fig. 1). The structure is described by the
1D Bogoliubov-de Gennes equation \cite{dG}:

\begin{equation}
\left\{\left[\left({ \hat p^2\over
2m}\right)-\tilde\mu+V\theta(L/2-|x|)\right]
\sigma_z + \hat\Delta (x)
\right\}\Psi=E\Psi,
\end{equation}
with the order parameter matrix given by

\begin{equation}
\hat\Delta(x)=\left(
\begin{array}{cc}
0 & \Delta e^{\displaystyle i\mbox{sign}x\phi/2}\\
\Delta e^{\displaystyle -i\mbox{sign}x\phi/2} & 0\\
\end{array}
\right)
\theta (|x|-L/2),
\end{equation}
$\tilde\mu=\mu-E_\perp$
is the chemical potential shifted by the energy of the transverse
mode, $\hbar=1$.

Consider first an isolated SI surface, assuming in Eq. (1)
$L=\infty$.
Constructing the ansatz for $|E|<\Delta$ as a superposition of
eigenfunctions,
with complex wave vectors
$k_\pm=\sqrt{2m(\tilde\mu \pm i\zeta)}$ $(\zeta=\sqrt{\Delta^2-
E^2})$
in the superconducting region and with decay rates
$\kappa_\pm=\sqrt{2m(V-(\tilde\mu\pm E))}$ in the insulating
region, we find
the dispersion equation,
keeping the difference between electron and hole wave vectors to
first
order,
\begin{equation}
{\zeta\over E}=-
{\delta\kappa k\over \kappa^2+k^2}.
\end{equation}
The surface level in Eq.(3) crucially depends on
the finite difference $\delta\kappa=\kappa_+ -\kappa_-$ of the
decay rates
of electron and hole wave functions inside the insulator,
while dephasing inside the superconductor is not important and
gives small
corrections omitted in Eq. (3).
The level lies close to the gap edge within the superconducting gap;
assuming a typical relation among the energies of the problem:
$\Delta\ll V-\tilde\mu\leq \tilde\mu$, one has
$\Delta-E\approx (\Delta^3/2V^2)( k/\kappa)^2$.
The wave function of the surface state decays
into superconductor on the characteristic length scale
$l\sim \xi_0\Delta/\sqrt{\Delta^2-E^2}$
($\xi_0$ is the coherence length) and into insulator on the
characteristic
length scale $l\sim 1/\kappa$.

Proceeding to calculation of the coupling of surface states at finite
$L<\infty$ in Eq. (1),
we construct wave functions by means of a
transfer matrix formalism \cite{Hu},
which yields the following dispersion relation:
\begin{equation}
E^2=\Delta^2\cos^2\left({|\beta|\pm\alpha\over 2}\right),
\;\;\; (|\beta| \pm \alpha>0),
\end{equation}
%
%
\begin{equation}
\cos\alpha= \tilde R + \tilde D\cos\phi,\;\;\;
\sin\beta= \tilde D\,\mbox{Im}(a_+a_-^\ast),
\end{equation}
where $\tilde D= \sqrt{D_+D_-},\;\; \tilde R=\sqrt{R_+R_-}$,
and the transmission  $D_\pm=|a_\pm|^{-2}$ and reflection
$R_\pm=1-D_\pm$ coefficients of normal electrons with energy
$\pm E$ are given in terms of the inverse transmission  amplitudes
\begin{equation}
a_\pm=\cosh\kappa_\pm L -
{i\over2}\left({\kappa_\pm\over k}-
{k\over\kappa_\pm}\right) \sinh\kappa_\pm L.
\end{equation}
Equation (4) is valid for all values of the barrier length $L$ and
barrier
heights $V$ \cite{com0}. A numerical solution of this equation is
presented in Fig. 2,
which also shows the corresponding wave functions.
To get an
explicit expression for the level energy we consider barriers which
are long
on the scale of quasiparticle decay $\kappa^{-1}$
but short on the scale of quasiparticle dephasing $|\delta\kappa|^{-
1}$:
$\;\;\kappa^{-1} \ll L \ll |\delta\kappa|^{-1}$ (the last condition is
equivalent to $(L/\xi_0)(k/\kappa)\ll 1$).
In this case, the junction transparency $\tilde D$, the dephasing
angle $\beta$ and
the coupling factor $\alpha$ are all
small: $\tilde D\ll 1$,$\;\;\;D_+ - D_-\approx -2\tilde D\delta\kappa
L$,
\begin{equation}
\beta\approx - {2k\delta\kappa \over
k^2+\kappa^2},\;\;\; |\beta|\ll 1,
\end{equation}
\begin{equation}
\alpha\approx 2\sqrt{\tilde D}
\sqrt{\sin^2(\phi/2)+(\delta\kappa L/2)^2}\ll 1.
\end{equation}
The last term in Eq. (8) is essential only in the close vicinity of
$\phi=0$:  $|\phi|\sim |\delta\kappa|L$.
Since the energy dispersion of $\beta$ can be neglected, one finds
from Eq. (4):
\begin{equation}
E_\pm^2=\Delta^2\left[1-\left({|\beta|\over 2}\pm
\sqrt{D}\left|\sin{\phi\over
2}\right|\right)^2
\right],
\end{equation}
where
$|\beta|/2\pm \sqrt{D}|\sin(\phi/2)|>0,\; D\approx \tilde D,\;
|\phi|\gg |\delta\kappa|L$.

The solution presented in Eqs. (4),(9) generally consists of
two pairs of bands $E(\phi)$: $\pm E_+(\phi)$ and $\pm E_-(\phi)$,
lying inside the superconducting gap
symmetrically with respect to the chemical potential.
The dephasing angle $\beta$ determines the position of the
preexisting surface states at
the single SI interface (cf. Eq. (3)), and the amplitude $\sqrt{D}$
determines the level splitting due to coupling via tunneling.
All branches of the bound state spectrum are fully developed if
$|\beta|/2>\sqrt{D}$ or
$ \kappa L> \ln|\kappa/\delta\kappa|$
\cite{com1}.

The dispersion of the bands $E(\phi)$ determines the current
flowing through
the bound states:
$I=2e(dE/d\phi)$ \cite{An,Be}.
Taking the derivative of Eq. (4), one gets under conditions (7),(8) the
current of
the single level in the form:
%
\begin{eqnarray}
I(E_\pm)=-\mbox{sign}E \;\; {e\Delta D \over 2} \sin\phi
\;\;\;\;\;\;\;\;\;\;\;\;\;\;\;\;\;\;\;\;\;\;\;\;\;\;\;\;\;\;\;
\nonumber\\
\;\;\;\;\;\;\;\;\;\;\;\;\;
\times \left( 1\pm {|\beta|\over 2\sqrt{D}}
{1\over\sqrt{\sin^2(\phi/2) + (\delta\kappa L/2)^2}}
\right).
\end{eqnarray}
%
%
The first term in Eq. (10) is consistent with calculations within the
tunnel
model, coinciding with the Ambegaokar-Baratoff current of a
single-mode junction
at $T=0$ \cite{AB}.
The second term dominates at small transparency $\sqrt{D}\ll\beta/2$.
It possesses an anomalous square-root dependence on the
junction transparency and anomalous (close to $\cos(\phi/2)$)
dependence on the phase difference, (Fig. 3). This current
corresponds to the transition of Cooper pair between
electrodes with the probability proportional to
the amplitude of the normal electron transition.
{}From a physical point of view, this  enhancement of the tunnel
current results from
resonance coupling of the surface states situated at SI interfaces of
the junction, Eq. (3).
This resonance coupling causes
large dispersion of the bound state bands, in analogy with the energy
level
splitting in the Schr\"{o}dinger symmetric double well potential
\cite{LL}.
We stress that
the resonant enhancement concerns only the tunneling of Cooper
pairs,
while the normal electron tunneling is non-resonant.

Thus, the bound levels
in low-transmission tunnel junctions carry a current which
tremendously exceeds the known critical Josephson
current.
However, this current is not manifested in equilibrium. Indeed,
assuming Fermi distribution $n_F(E_\pm)$ for the occupation
numbers of the
bound levels, one finds with the assumed accuracy the current of
all bound states:
%
\begin{eqnarray}
I_{bound}=\sum_{signE,\pm} I(E_\pm) n_F(E_\pm)
\;\;\;\;\;\;\;\;\;\;\;\;\;\;\;\;\;\;\;\;\;\;\;\;\;
\nonumber\\
\;\;\;\;\;\;\;\;\;\;\;\;\;\;\;\;\;\;\;\;\;\;\;\;\;\;
=e\Delta D\tanh(\Delta/2T)\sin\phi.
\end{eqnarray}
%
This current is twice bigger than the magnitude of the Josephson
current
calculated for a single-mode junction within the tunnel model
\cite{AB} and
beyond it \cite{Hab}.
However, in order to get the total current one has to take
into account the contribution from the continuum:
\begin{equation}
I_{cont}={e\over\pi}\int dE
[(|t^N(\phi)|^2-|t^A(\phi)|^2)- (\phi\rightarrow -\phi)]
n_F(E).
\end{equation}
The normal transmission amplitude $t^N$ in Eq. (12) has the form:
\begin{equation}
t^{N}=\tilde D{
E[(E-\xi)a_+e^{i\phi/2}-(E+\xi)a_-e^{-i\phi/2}]
\over
\Delta^2(\cos\alpha+\cos\beta)-2E^2\cos\beta+2i\xi E\sin\beta},
\end{equation}
where $\xi=\sqrt{E^2-\Delta^2}$. The Andreev transmission
amplitude $t^A$ is
symmetric in $\phi$ and drops out of Eq. (12), which together with
Eqs. (7),(8),(13), yields:
\begin{equation}
I_{cont}= - {e\Delta \over 2} D\tanh(\Delta/2T)\sin\phi
= -{1\over 2}I_{bound}.
\end{equation}

The calculation presented above, as well as the results of direct
numerical calculations in Fig. 3,
uncover a remarkable fact,
namely that the Josephson
current in a symmetric junction with an extended tunnel barrier
results from cancellation of large
resonant currents flowing through individual bound states. This
situation
resembles the situation in a normal junction: while individual
scattering states of electrons carry a finite current, the total
current
through the junction is equal to zero in equilibrium due to
cancellation of
currents of the modes incident from the right and from the left.
To reveal the net current of a single mode, one has to create a
current imbalance, connecting the junction to reservoirs with
different
chemical potentials. In a similar way, creation of current imbalance
in the Josephson junction by
means of nonequilibrium population of the bound states is able to
reveal the
current of a single bound state. For example,
it is possible to equalize the level populations within
one of the bound level pairs (e.g. $\pm E_-$) by means of resonant
electromagnetic pumping, as suggested
in Ref.\cite{Sh}. In principle, one might then suppress the current of
this level pair and thereby
reveal the current of the second pair of levels ($\pm E_+$),
which will show up in a Josephson current enhanced by the factor of
$|\beta|/2\sqrt{D}$, flowing in the same direction
as the equilibrium current.
Suppression of the current of the other level pair $\pm E_+$, by
a proper choice
of the frequency of the pumping field, will show up in
an enhanced current flowing in the opposite direction.
Another possibility is to inject excess quasiparticles into one of the
bound levels by means of tunnel coupling to an additional normal
electrode, i.e. to use a three-terminal device similar to the one
suggested by
van Wees et al. \cite{Ha} in order to reveal the currents of
individual Andreev states in
a SNS junction (see also \cite{Kl}).

A crucial condition for observing the current
of a single bound state in experiment is sufficient resolution of the
bound
level structure.
Since the major part of the bound state wave function is
concentrated in
the electrodes, the main mechanism of level broadening is
quasiparticle recombination due to electron-phonon scattering in
the superconducting banks.
According to Kaplan et al.
\cite{Ka} this gives an estimate for the level width:
$\Gamma/\Delta \sim 10^{-2}$ at the critical
temperature, decreasing exponentially at low temperature.
The interlevel distance in the optimal case of a single-mode
junction, according to
Eq. (10), is $\delta E\approx \Delta\beta\sqrt{D}$, decreasing in
multimode
junctions inversely proportionally to the number of transverse
modes.
Combination of the resolution condition $\delta E>\Gamma$ with
the condition
of a long junction, $\beta>2\sqrt{D}$,
yields for single-mode junction $\beta^2>2\Gamma/\Delta$.
The factor $\beta^2$ depends on the
height of the tunnel barrier and varies between
$(\Delta/\mu)^2<\beta^2<\Delta/\mu$ in junctions with high
($V\sim\mu$) and
low ($V-\mu\sim\Delta$) tunnel barriers.
In line with these estimates, one might expect to observe the effect
- giant Josephson current, Eq. (10) - in
junctions with a very low tunnel barrier, like
gate controlled S-2DEG-S devices, or with large ratio $\Delta/\mu$,
like high $T_c$ junctions.

A more realistic approach, however,
would be to increase the energy level splittings, increasing the
$\beta$ factor by making a SNI well at the
interface \cite{Blamire}. For a symmetric SNINS structure we find
the same result for the bound state spectrum as Eq.(4), but
with a factor $\beta=4d/\xi_0$, $\;\; d\ll \xi_0$,
where $d$ is the width of normal region, corresponding
to preexisting surface state in SNI well \cite{dG2}.
This can easily provide a
sufficiently large $\beta$-factor, say  $\beta\sim 0.1$ \cite{com2}.

In conclusion, we have studied a microscopic mechanism of
Josephson current
transport in single-mode tunnel junctions with wide quasiclassical
tunnel  barriers.
We found that each Andreev bound state in such junctions carries a
giant current proportional to the amplitude of normal electron
transmission through the junction.
Experimental observation of the current of a single bound state is
possible under nonequilibrium conditions.

We gratefully acknowledge discussion with A. Slutskin and R.
Shekhter. One of the authors (GW) acknowledges computational
assistance by A. Berntsson during the initial stages of this work.
This work has been supported by the
Swedish Natural Science Research Council (NFR), and by the NUTEK/NFR
Interdisciplinary Consortium on Superconducting Materials.

\begin{figure}
\label{Fig1}
\caption{Single-transverse-mode adiabatic junction modeled
as a 1-dim SIS junction
with a wide barrier of length $L$ and height $V$.}
\end{figure}

\begin{figure}
\label{Fig2}
\caption{ Andreev level energies and wave functions in a single-mode
tunnel junction.
Left part: Andreev level energies $E(\phi)>0$ as functions of phase
difference $\phi$ across the junction.
Full lines: Long SIS junction ($L=1$, $V=50$);
Dashed line: Short SIS junction ($L=0.2$, $V=270$) having the same
transparancy $D$ as the long junction but showing only a single
Andreev level, $E_+$ [12].
Right part: Andreev wave functions
$|\psi|^2 = |u|^2+|v|^2$ for the long SIS junction at $\phi=\pi$;
each of the $u$ and $v$ components are
symmetric or antisymmetric around $x=0$, making
$E_\pm$ nearly "bonding" or "antibonding" states.
In this figure the choice of parameters is beyond the approximate
Eqs. (8),(9) to achieve suitable resolution.
}
\end{figure}

\begin{figure}
\label{Fig3}
\caption{
DC Josephson currents in a single-mode tunnel junction at $T=0$
(long SIS junction:
$D \approx \; 2.4\times10^{-9}$, $\beta \approx \; 3\times10^{-2}$).
$I_+=I(-E_+)$ and $I_-=I(-E_-)$ are giant Josephson currents
associated with individual Andreev levels.
$I_{bound} =I_+ + I_-$ is the small residual current from
compensating giant bound state currents. $I_{cont}$ is the
continuum contribution to the current (Eq. (12)). $I_J=I_{bound} +
I_{cont}$ is the (total) Josephson current of the junction.
}
\end{figure}


\begin{thebibliography}{99}
\bibitem{com0} Eq. (4) allows also analytical continuation to the
region $V<\mu$ which corresponds to a SNS junction; it therefore
gives a
unified description of SIS and SNS junctions.
\bibitem{com2} The restriction on the possible non-asymmetry is
$|d_1-d_2|\ll\xi_0\sqrt{D}$.
\end{thebibliography}
\end{document}